\documentclass{article}

\usepackage{arxiv}

\usepackage[utf8]{inputenc} 
\usepackage[T1]{fontenc}    
\usepackage{hyperref}       
\usepackage{url}            
\usepackage{booktabs}       
\usepackage{amsfonts}       
\usepackage{nicefrac}       
\usepackage{microtype}      
\usepackage{lipsum}
\usepackage{graphicx}
\usepackage{float}

\title{Software Framework for Tribotronic Systems}

\author{
  Jarno Kansanaho\thanks{Corresponding author} \\
  Faculty of Information Technology\\
  University of Jyv{\"a}skyl{\"a}, FI-40014, Finland\\
  \texttt{jarno.m.kansanaho@jyu.fi} \\
   \And
 Tommi K{\"a}rkk{\"a}inen \\
  Faculty of Information Technology\\
  University of Jyv{\"a}skyl{\"a}, FI-40014, Finland\\
  \texttt{tommi.karkkainen@jyu.fi} \\
}

\begin{document}
\maketitle

\begin{abstract}
Increasing the capabilities of sensors and computer algorithms produces a need for structural support that would solve recurring problems. Autonomous tribotronic systems self-regulate based on feedback acquired from interacting surfaces in relative motion. This paper describes a software framework for tribotronic systems. An example of such an application is a rolling element bearing (REB) installation with a vibration sensor. The presented plug-in framework offers functionalities for vibration data management, feature extraction, fault detection, and remaining useful life (RUL) estimation. The framework was tested using bearing vibration data acquired from NASA's prognostics data repository, and the evaluation included a run-through from feature extraction to fault detection to remaining useful life estimation. The plug-in implementations are easy to update and new implementations are easily deployable, even in run-time. The proposed software framework improves the performance, efficiency, and reliability of a tribotronic system. In addition, the framework facilitates the evaluation of the configuration complexity of the plug-in implementation.
\end{abstract}

\keywords{Software framework \and Tribotronic system \and Bearing diagnostics \and Bearing prognostics \and Vibration analysis}

\section{Introduction}
The term 'tribology' was introduced and defined in The Jost Report  \cite{JOSTREPORT1966} as "the science and technology of interacting surfaces in relative motion and of the practices related hereto." It was reported that enormous amounts of resources were wasted because mechanical surface phenomena was ignored \cite{BROSTOW2013}. However, The Jost Report did not pay much attention to wear, the most significant tribological phenomenon \cite{BROSTOW2013}. Tribology enables the effective design of both machines and lubrication to minimize the impact of friction and wear \cite{GLAVATSKIH2008_TRIBO}. The successful implementation of tribological practices into design procedures for various machines and mechanisms has resulted in significant economic savings through improvements in machine performance and reliability \cite{GLAVATSKIH2008_TRIBO}. Tribology combines physics, chemistry, materials engineering, machinery theory, and products of its own engineering science \cite{KIVIOJA2004_TRIBO}. Holmberg \cite{HOLMBERG2001_TRIBO} defined a taxonomy for different levels of occurrence of tribological phenomena: universe, global, national, plant, machinery, component, contact, asperity, and molecular. Each level of this classification comprises its own components and interactions between them.

A system is a set of related and interdependent elements that regularly interact to form an integrated whole \cite{BACKLUND2000}. At high levels of abstraction, a tribological system can be described with input and output variables and the interaction between these variables. The input variable is energy (e.g. force, moment and kinematics) and the output variables are matter and signals \cite{KIVIOJA2004_TRIBO}. The interaction between elements causes friction and wear losses that are summarized as loss-outputs \cite{CZICHOS2000:_TRIBO}. A tribosystem is a tribological system that includes at least two contacting tribological components \cite{BLAU2016_TRIBOSYSTEM}. A number of input and output variables in tribosystems can be infinite due to the number of physical and chemical properties of the surfaces in contact, the properties of the medium (i.e., the lubricant), and the environmental conditions \cite{KIVIOJA2004_TRIBO}. For example, bearings, gears, and mechanical seals are tribological components. Glavatskih et al. \cite{GLAVATSKIH2008_TRIBO} outlined a tribotronic system that would unite the tribosystem, sensors, real-time control system, and actuators. The tribotronic system is distinct from a mechatronics system because tribosystems use loss outputs, such as wear, vibration, temperature, and friction. Controlling these outputs allows a tribosystem to try to improve the performance, efficiency, and reliability of the whole machine \cite{GLAVATSKIH2008_TRIBO}. A desirable output for a tribotronic system should be expressed in terms of endurance life and probability of failure \cite{HOLMBERG2001_TRIBO}.

One of the main goals of software engineering is to reuse existing code \cite{BOSCH2000:_FRAMEWORK}. A software's framework is a "skeleton" that can be used to supplement application-specific software, so recycling existing frameworks is a key technique when implementing software platforms. Software frameworks are customized to complete a software application by filling empty code blocks with product-specific code. An important property of software frameworks is inversion control, which enables the framework itself to call user-implemented methods, that is not possible in traditional procedural programming \cite{JOHNSON1988_FRAMEWORK}. If frameworks were not reused during software development, a considerable amount of code would be written repeatedly. Our study focused on object-oriented frameworks. Abstract frameworks provide only software interfaces; they do not include any runnable code. White-box frameworks use subclasses as extensions, which allow the implementation of methods for base classes. Black-box frameworks use a composition approach and include ready-to-use classes. It should be noted that white-box frameworks evolve into black-box frameworks over time \cite{JOHNSON1988_FRAMEWORK}. Gray-box frameworks merge black-box and white-box issues \cite{MATTSSON_1999}. Layered frameworks can be applied to large-scale platforms when different frameworks need to be fused \cite{KAISLER2005:_FRAMEWORK}. Plug-in frameworks are specialized because they implement application-specific interfaces, or plug-ins \cite{HERMANN2012:_FRAMEWORK}. Caropreso et al. \cite{CAROPRESO2019_FRAMEWORK} presented a structured methodology to define the architecture for communication frameworks with multiframe capabilities. It is an example of maintainable object-oriented framework that is applicable for embedded systems.

\begin{figure}[h]\centering
	\includegraphics[width=0.8\linewidth]{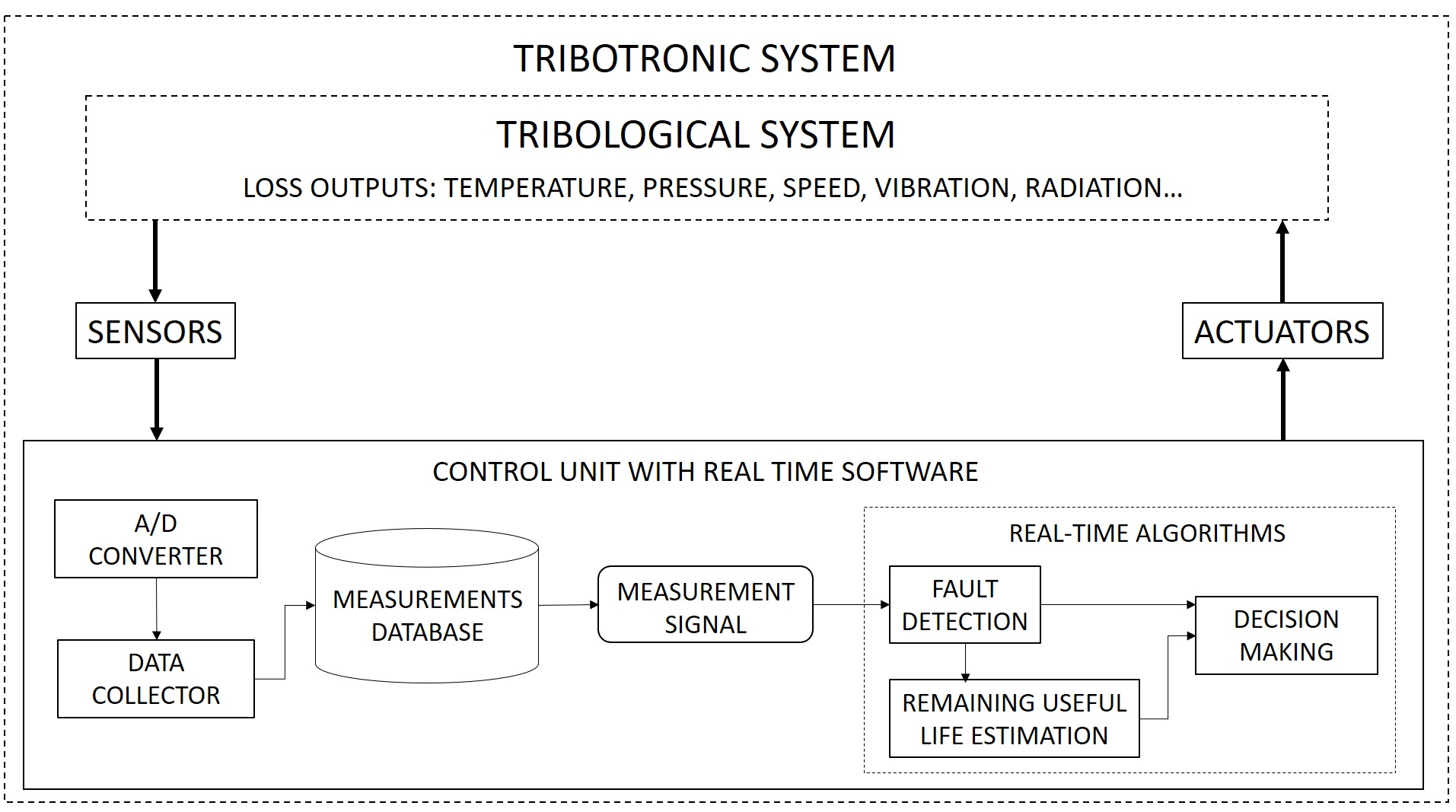}
	\caption{Tribotronic system \cite{GLAVATSKIH2008_TRIBO}}\label{FIG_1}
\end{figure}

Figure \ref{FIG_1} depicts a schematic tribotronic system with a control unit and real-time software. In this paper, we introduce an object-oriented plug-in framework for such tribotronic systems. The main motivation for this work is to speed up and ease the deployment of diagnostic and prognostic algorithms into tribotronic systems. The purpose of the framework is to improve the performance, efficiency, and reliability of a tribotronic system. The framework covers asset and data management, fault detection, and RUL estimation. The plug-in implementation was targeted for REBs that were monitored using vibration sensors.

The contents of this article are as follows. Bearings (tribosystem) and their wear evolution will be introduced in Section \ref{sec2}. Vibration analysis, fault detection, and RUL estimation will be explained in Section \ref{sec3}. Designed plug-in framework will be presented in Section \ref{sec4}. Evaluation of the framework will be presented in Section \ref{sec5}. Finally, conclusions and future work will be summarized in Section \ref{sec6}.

\section{Tribosystem - Bearing}\label{sec2}

Bearings are widely used in rotating machinery to support shafts. Bearings are categorized as either REBs or journal bearings based on their structure \cite{HAMROCK1983:_REBS}. REBs contain spherical, cylindrical, tapered, and needle-shaped rolling elements. Journal bearings contain only sliding surfaces --no rolling elements. Monitoring the condition of these bearings is very important because a bearing failure is a very common reason for machine breakdowns. In general, the vibration and temperature of a tribological system (REB) are monitored to detect lost outputs.

\subsection{Bearing failure}

Bearing failures fall under six categories: fatigue, wear, corrosion, electrical erosion, plastic deformation, and fracture/cracking \cite{SKF2012:_REB_BOOK}. Wear is a cumulative quantity regularly measured by condition monitoring systems \cite{CHRISTER1995_WEAR}. When a measured variable directly determines a bearing's failure, the condition monitoring method is direct; when a measured variable provides information associated with and affected by the bearing's condition, the condition monitoring method is indirect \cite{CHRISTER1995_WEAR}. Common direct and indirect condition monitoring methods consist of the following \cite{DRAGAN2008_CBM,JANTUNEN2002_CBM,WANG2017_VIB,DONGRE2013_WEAR}: i) indirect methods include monitoring vibrations, acoustic emissions, basic physical quantities such as heat and pressure, basic electrical quantities such as voltage, current, power, and resistance, and ultrasound or infrared testing, and ii) direct methods include oil debris or corrosion analysis as well as visual inspection using a borescope. Furthermore, new methods are constantly being sought that would be more sensitive when measuring bearing defects \cite{RAI2016_REVIEW_VIB_DIAG}.

\begin{figure}[h]
\centering
	\includegraphics[width=0.8\linewidth]{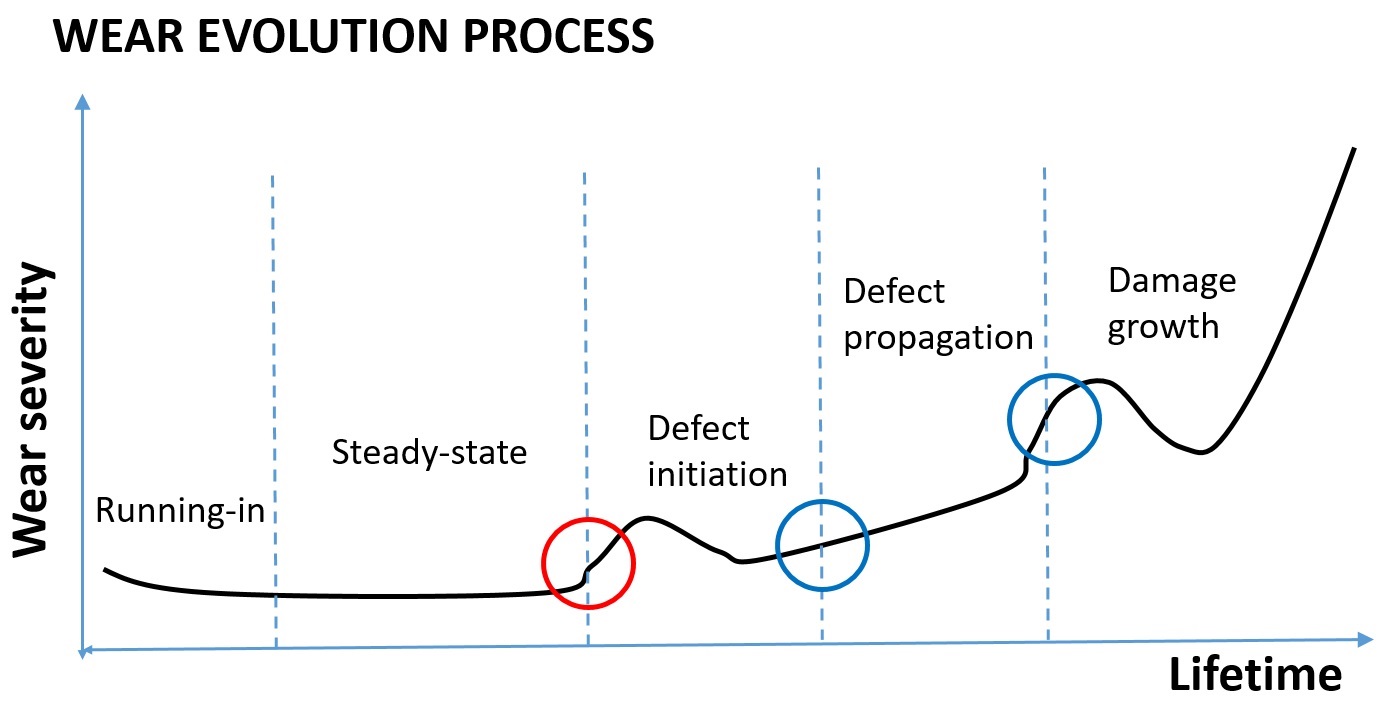}
	\caption{Wear evolution process \cite{IDRISS2014_WEAR}}\label{FIG_2}
\end{figure}

Presenting wear evolution of REBs as a time series describes the wear interaction and evolution at different lifetime stages. A five-stage descriptive model of lifetime stages, as depicted in \ref{FIG_2}, was presented by El-Thalji et al. \cite{IDRISS2014_WEAR}: running-in, steady-state, defect initiation, defect propagation, and damage growth. First, during the running-in stage, the surface asperities and the lubrication film become uniform \cite{IDRISS2014_WEAR}. The length of the steady-state stage, the healthy stage of the lifetime, depends on maximum load-induced stress, material characteristics, and operating temperature \cite{ARAKERE2012_WEAR}. The wear process starts and will affect surface roughness and waviness in the defect initiation stage \cite{COOKSON2011_WEAR}. According to \cite{IDRISS2014_WEAR}, this stage can be further split into the sub-stages of defect localization, dentation, crack initiation, and crack opening. The linear elastic fracture mechanics commences in the defect propagation stage \cite{PANASYUK1995_WEAR}. Incubation, stable, and crack-to-surface are then the main events that occur \cite{OLVER2005_WEAR}. The defect starts to grow in three dimensions (length, width, and depth) and the effect of multiple asperities is prominent in the damage growth stage \cite{IDRISS2014_WEAR}. Direct condition monitoring makes it possible to detect lifetime stages of wear evolution directly from measurements; however, this is not possible using raw vibration measurements.

\section{Vibration analysis}\label{sec3}

Vibration sensors interpret vibration values indirectly using mechanical and optical quantities. Vibration sensors are categorized as contacting or non-contacting according to their measurement principles. Both contacting and non-contacting sensors are further divided according to path, speed, and acceleration measurement. Path measurement uses potentiometric transmitters and linear variable differential transformers; speed is measured using principles of electrodynamics and seismometers; acceleration is measured using piezoelectric, piezo-resistive, resistive, and inductive sensors \cite{RUHM2010:_SENSORS}.

A machine's vibrational signature is related to either a standard condition or a fault condition \cite{RANDALL2010_VIB_CM}. A tribotronic system measures vibrations and processes them to discover informative features using feature extraction. Frequently, features calculated from vibration signals are high-dimensional and non-Gaussian. Further, feature selection is applied to extracted features to leave over the most relevant features. Descriptive classification for features of vibration signals is the following: i) time-domain features; ii) frequency-domain features; iii) time-frequency-domain features; iv) phase-space dissimilarity measurements; v) complexity measurements; vi) other features \cite{CAESARENDRA2017_VIB}. A considerable amount of research has been directed towards the development of the digital signal processing of vibrational signals \cite{RAI2016_REVIEW_VIB_DIAG}.

\subsection{Fault detection}

Randall stated \cite{RANDALL2010_VIB_CM} that "fault detection is the first step in the overall process of detection, diagnostics and prognostics. Since all signals have to be processed to determine whether a significant change has occurred, the techniques employed must be considerably more efficient than those which might be used for the latter processes." Early fault detection allows time to predict fault progression and estimate RUL before catastrophic failures occur \cite{ZHANG2011_FAULT_DETECT}. Fault detection is one of the main functionalities in the designed framework. Depending on the response from the fault detection algorithm, the tribosystem (REB) would be controlled by actuators.

\subsection{Remaining useful life estimation}

RUL estimation is an important prognostic and health management task that enables optimized maintenance plans to enhance production, minimize costly downtime, and avoid catastrophic breakdowns \cite{WEI2014_RUL, WANG2016_RUL_PF}. RUL estimation approaches are categorized into physical model approaches, data-driven approaches, and hybrid approaches \cite{MEDJAHER2012_RUL, WEI2014_RUL}. Further, the data-driven approaches can be categorized into knowledge-based, statistical, and supervised methods \cite{SIKORSKA2011_RUL_REVIEW}. Recent machine learning approaches have frequently been applied to the diagnoses and prognoses of REBs \cite{TRAN2019_BIGDATA_FEAT}. However, the effectiveness of the machine learning methods rely on the quality of features of vibration signals.

An ideal signal processing method should be capable of detecting the bearing degradation phases on changing defect conditions \cite{RAI2016_REVIEW_VIB_DIAG}. Crucial for RUL estimation is to find the most suitable feature to describe the degradation process when vibration measurements are used. Measuring vibrations is an indirect methods to monitor the condition of REBs. RUL estimation is another main functionality of the framework.

\section{Implementation of the framework}\label{sec4}

\begin{figure}[h]
\centering
	\includegraphics[width=\linewidth]{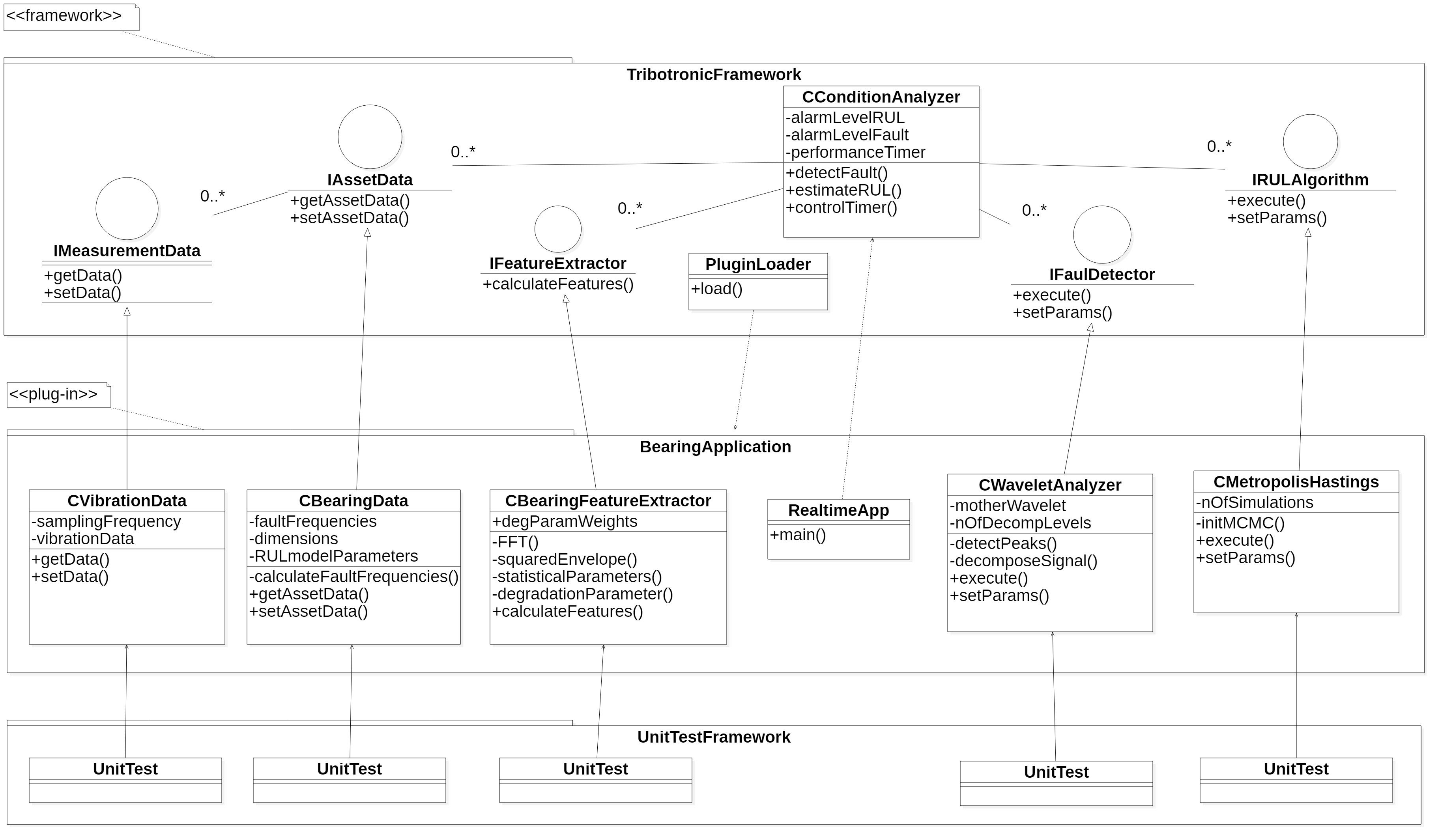}
	\caption{Tribotronic plug-in framework}\label{FIG_3}
\end{figure}

The plug-in framework designed for tribotronic systems supports asset and data management, feature extraction, fault detection, and RUL estimation. The framework design is shown in Figure \ref{FIG_3}. A measurement database can be deployed in a local or remote computer. The results of the fault detection and RUL estimation are inputs for a condition analyzer that passes the results to a module that decides how the tribosystem is controlled. The general architecture of the system has been presented Figure \ref{FIG_1}. A component linking the condition analyzer, the decision-maker, and the actual control of the system was not explicitly defined in the framework because it would depend on the machine, according to the sorts of actions needed to maintain running conditions or stop the machine's operation.

The framework includes interfaces for measurement data (\texttt{IMeasurementData}), asset data (\texttt{IAssetData}), feature extraction (\texttt{IFeatureExtractor}), fault detection (\texttt{IFaultDetector}), and RUL algorithms (\texttt{IRULalgorithm}). The measurement data is acquired from the sensor and the asset data is relates to the tribotronic component in question. Fault detection and RUL estimation are executed by the \texttt{CConditionAnalyzer} class. The \texttt{PluginLoader} class loads the desired plug-in that includes the appropriate implementations for the application in question. The \texttt{BearingApplication} plug-in implements the interfaces of the framework using inheritance (Figure 3). The presented plug-in was designed for REBs. The plug-in implementations are based on previous research on feature extraction from vibration signals, fault detection, and RUL estimation of REBs.

Technical bearing data is included in the \texttt{CBearingData} class. The bearing dimensions are the minimum data required to calculate characteristic fault frequencies, which are required for the fault detection algorithm. Measurement data is specialized as a \texttt{CVibrationData} class that handles vibration data. Vibration signals are loaded from the database as shown in Figure \ref{FIG_1}. Vibration data includes vibration signals and their measurement time and sampling frequency. Further, specialization of the measurement data could be easily done, for example, to address temperature and oil debris data that represent other commonly used condition monitoring measurements for REBs.

The \texttt{CConditionAnalyzer} class uses the \texttt{CBearingFeatureExtractor} that includes methods to extract specified features from vibrational signals. Methods used to calculate statistical features, such as RMS, skewness, and Kurtosis, include an optimal degradation parameter, a fast Fourier transform (FFT), and a squared envelope spectrum. The FFT is a sub-routine for the squared envelope spectrum calculation. Previously, an introduced fault detection algorithm was implemented into \texttt{CWaveletAnalyzer}.

The fault detection algorithm is called by \texttt{CConditionAnalyzer} and implemented in the \texttt{CMetropolisHastings} class. The algorithm requires the characteristic fault frequencies of an REB and the sampling frequency of a vibrational signal as input parameters. The algorithm returns a boolean value indication of a fault or not-fault status.

Similar to fault detection, the RUL estimation algorithm is called by \texttt{CConditionAnalyzer}. The model parameters are stored in the \texttt{CBearingData} class when a model-based RUL estimation is applied. The accuracy calculations, as defined in equation \ref{eq:1}ref{eq:1}, which were used to determine degradation features were implemented in the \texttt{CBearingFeatureExtractor} class. The best degradation feature, the RUL model parameters, and the alarm level are input parameters for the RUL estimation algorithm that was implemented in the \texttt{CMetropolisHastings} class. The RUL algorithm returns the time of the last operation date.

The plug-in implementations are tested with unit tests. Unit tests are carefully designed to test the smallest components. Unclear definitions of unit testing leads to bad and inconsistent testing and makes the software error-prone \cite{RUNESON2006:_UNIT_TESTING}. This guarantees the reliability of the framework being run and can be updated in real-time.

\begin{figure}[h]
\centering
	\includegraphics[width=\linewidth]{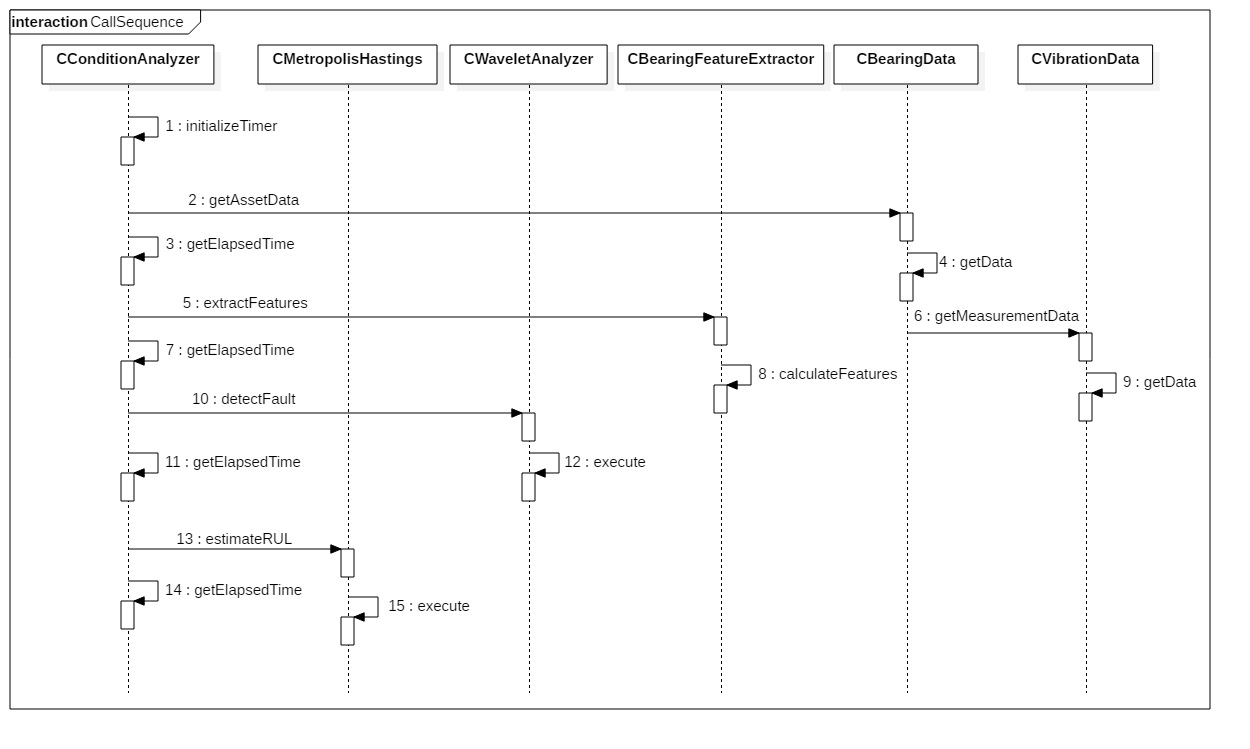}
	\caption{Sequence diagram of fault detection and RUL estimation}\label{FIG_4}
\end{figure}

A sequence diagram of fault detection and RUL estimation with the suggested realization of the framework are shown in Figure \ref{FIG_4}. The performance of the framework is measured by an internal timer initialized in the first call. Elapsed times for data loading, feature extraction, fault detection, and RUL estimation are recorded. However, RUL estimation is not executed if the result from fault detection is negative, indicating a non-faulty bearing.

A very important aspect of a system is its configuration complexity \cite{KELLER2007:_CONFIG}. A complex algorithm does not need to be complex to configure; e.g. \cite{EARLYSPD2018:_FAULT_DETECT_PATENT}. Complicated configurations can be error-prone and time-consuming, which increases the cost of the system. The meta-parameters of the algorithms play the key role in the evaluation of the configuration complexity. The configuration complexity can be evaluated based on the meta-parameters in the plug-in implementation: \textit{alarmLevelFault},  \textit{motherWavelet}, \textit{nOfDecompLevels}, \textit{degParamWeights}, \textit{alarmLevelRUL}, \textit{RULmodelParameters} and \textit{nOfSimulations}.

\subsection{Specific algorithm descriptions}

The plug-in's implementation of the fault detection algorithm uses time-frequency domain features. The algorithm exploits discrete spline wavelet decomposition with bior6.8 as the basis wavelet \cite{KANSANAHO2018_DIAG_WAVELET}. The squared envelope spectrum of the reconstructed signals is searched for characteristic fault frequencies for each wavelet decomposition level, and the peaks are detected based on local maxima. The peak detection algorithm uses a user-defined alarm level.

A method proposed by Zhang et.al. \cite{ZHANG2016_RUL} for degradation feature selection was integrated into a plug-in; the method defines the feature goodness metrics of correlation, monotonicity, and robustness. The optimal degradation feature is selected using a weighted linear combination of the proposed metrics:
\begin{equation}
\mathbf{\max_{X \in \Omega}} J = \omega_1Corr(X) + \omega_2Mon(X) + \omega_3Rob(X), \label{eq:1}
\end{equation}
where J is the score value, $\Omega$ is the set of candidate degradation features, and $\omega_i$ is the weight for individual metrics. 

The implemented RUL estimation algorithm is based on the adaptive Metropolis-Hastings algorithm so it can calculate the parameters for the degradation model. The Metropolis-Hastings algorithm is a sampling algorithm based on Markov-Chain-Monte-Carlo (MCMC) algorithm \cite{ROBERTS2004:_MCMC}. MCMC methods aim to solve multi-dimensional integrals using numerical approximations. The Metropolis-Hastings algorithm generates a random walk using a proposal density and a method for rejecting some of the proposed moves. In our study, an Adaptive Metropolis (AM) algorithm is used where the Gaussian proposal distribution is updated along the process using the complete information cumulated so far \cite{HAARIO2001_MCMC}. A simple exponential degradation model is used in the evaluation \cite{GOEBEL2008_RUL_EXP, AN2013_RUL_PF}:
\begin{equation}
\mathbf{Deg = c \exp(b t)} ,
\label{eq:2}
\end{equation}
where $c$ and $b$ are the model parameters, $t$ is the time, and $Deg$ is the degradation indicator. The exponential model is very often used in RUL estimation for REBs, although different modifications have been suggested \cite{LI2015_RUL_EXP}.

\begin{table}[h]
\caption{IMS BEARING DATA SPECIFICATIONS}
\centering
\label{table_1}
\centering
\begin{tabular}{ | p{9.0cm} | p{7.0cm} | }
\multicolumn{2}{l}{BEARING INFORMATION}\\
\hline
Bearing model & Rexnord ZA-2115 \\
\hline
Bearing type & Double row bearing \\
\hline
\multicolumn{1}{l}{}\\
\multicolumn{1}{l}{VIBRATION MEASUREMENTS}\\
\hline
Number of bearings & 4\\
Sampling frequency & 20480 Hz \cite{GOUSSEAU2016_IMS}\\
Sampling length & 20480 data points\\
Shaft rotation speed & 33.3 Hz\\
Inner race fault frequency (BPFI) & 297 Hz\\
Outer race fault frequency (BPFO) & 236 Hz\\
Rolling element fault frequency (BSF*2) & 278 Hz\\
\hline
DATASET 1 &\\
Number of measurements & 2156\\
Recording duration & 34 days 12 hours\\
Detected bearing faults & Bearing 3 / BPFI, Bearing 4 / BSF\\
\hline
DATASET 2 &\\
Number of measurements & 984\\
Recording duration & 6 days 20 hours\\
Detected bearing faults & Bearing 1 / BPFO\\
\hline
DATASET 3 &\\
Number of measurements & 4448\\
Recording duration & 31 days 10 hours\\
Detected bearing faults & Bearing 3 / BPFO\\
\hline
\end{tabular}
\end{table}

\section{Experimental results}\label{sec5}

A popular REB dataset from the Center for Intelligent Maintenance Systems (IMS) of the University of Cincinnati was used in the evaluation. Much research has analyzed the IMS dataset \cite{QIU2006_DIAG_WAVELET_IMS, MAHAMAD2010_RUL_IMS, SOUALHI2014_RUL_IMS, GAUTIER2015_DIAG_IMS, GOUSSEAU2016_IMS, ARUN2017_DIAG_IMS, AHMAD2018_RUL_IMS}. In these run-to-a-failure-tests, four Rexnord ZA-2115 double row bearings were installed on one shaft. The shaft rotation (2000 RPM) and the radial load (6000 LBS) were constant during the test-runs and all bearings were force-lubricated. During the test-runs, the designed lifetime of the bearing was exceeded for all failures. The IMS bearing data specifications are collected in Table \ref{table_1}.

Datasets for the evaluation were selected based on recent findings by \cite{GOUSSEAU2016_IMS}. Selected cases of faulty bearings include an inner fault in bearing 3 (Dataset 1) and an outer race fault in bearing 1 (Dataset 2). Vibration signals were processed through the realizations of the \texttt{IFilter} and \texttt{ITransform} interfaces, which provide feature extraction methods. The resulting features are shown in Table \ref{table_2}.

\begin{table}[h]
\caption{VIBRATION SIGNAL FEATURES}
\centering
\label{table_2}
\centering

\begin{tabular}{ | p{9.0cm} | p{7.0cm} | }
\multicolumn{2}{l}{STATISTICAL TIME DOMAIN FEATURES}\\
\hline
FEATURE & DESCRIPTION\\
1. Root Mean Square (RMS) & The power content\\
2. Crest Factor & The ratio of the peak amplitude to the RMS\\
3. Shape Factor & The RMS divided by the signal mean\\
4. Impulse Factor & The maximum of the peak amplitudes divided by the signal mean\\
5. Shannon Entropy & The degree of uncertainty\\
6. Log Energy Entropy & The degree of uncertainty\\
7. Skewness & Asymmetry measure of the PDF of the signal\\
8. Kurtosis & The impulsiveness of the signal\\
\hline
\end{tabular}

\begin{tabular}{ | p{16.5cm} | }
\multicolumn{1}{l}{FREQUENCY DOMAIN FEATURES}\\
\hline
Squared Envelope Spectrum\\
9. - Amplitude of characteristic defect frequency (1. harmonic)\\
\hline
\multicolumn{1}{l}{TIME-FREQUENCY DOMAIN FEATURES}\\
\hline
Re-constructed signal of Wavelet Decomposition Levels (bior6.8)\\
Squared Envelope Spectrum\\
10. - Amplitude of characteristic defect frequency (1. harmonic)\\
\hline
\end{tabular}

\end{table}

The fault detection algorithm calculates the squared envelope spectrum of a vibration signal. The amplitudes of the characteristic fault frequencies are identified from the envelope spectrum using peak detection. The fault state indication is determined using an alarm level three times the mean of the maximum amplitude of the envelope spectrum of the steady state (i.e., non-faulty) signal. The alarm level was justified earlier based on the noise level of non-faulty vibration signals.

\begin{figure}
    \centering
	\includegraphics[width=\linewidth]{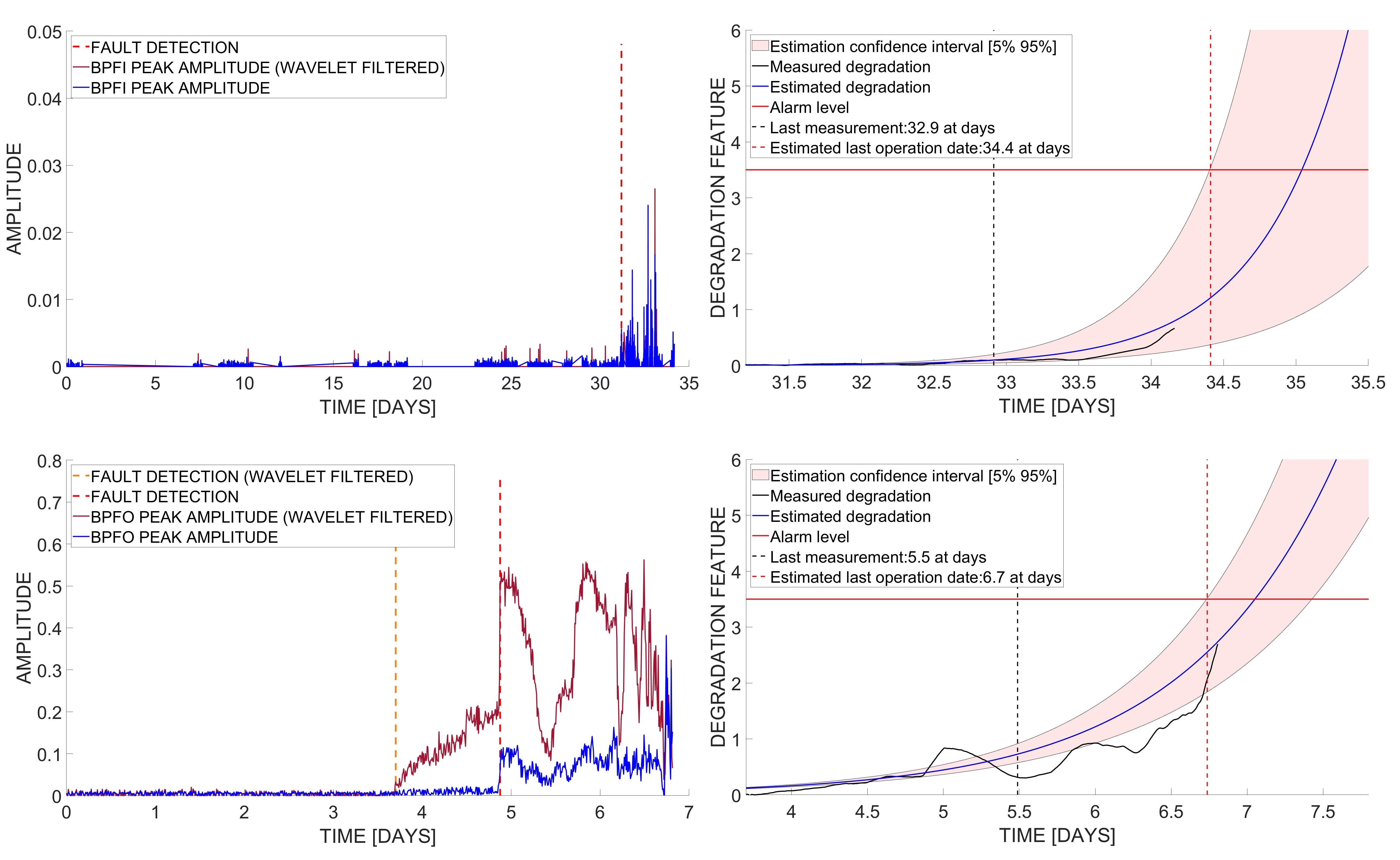}
	\caption{Fault detection (LEFT) and RUL estimation (RIGHT): Bearing 3(TOP), Bearing 1(BOTTOM)}\label{FIG_5}
\end{figure}

The left side of Figure \ref{FIG_5} illustrates the fault detection for both bearings, with bearing 3 on the top and bearing 1 on the bottom. The dotted red line represents the fault detection time. The blue line represents the amplitude of the ball pass frequency of the inner race (BPFI) in the envelope spectrum of the vibration signal. The orange plot is the highest amplitude of the BPFI in the envelope spectrum of the re-constructed signals. The BPFI was detected after testing bearing 3 for 31.5 days (Figure \ref{FIG_5}). The wavelet decomposition was determined for twelve levels, resulting in the highest frequency band from 5 kHz to 10 kHz. The wavelet filtering does not give any earlier indication of the fault compared to the envelope spectrum. The ball pass frequency of the outer race (BPFO) was detected at 3.8 days into the testing of bearing 1. The wavelet filtered signal provided fault detection 1.1 days earlier than the envelope spectrum.

RUL estimation was processed following fault detection using the adaptive Metropolis-Hastings MCMC algorithm. The best degradation feature was determined using the goodness metrics defined in equation \ref{eq:1}. Table \ref{table_3} includes the goodness metrics calculations for the selected features. All the features were normalized to the same sampling rate and the features were smoothened using a moving average with a window of 20 samples. In terms of time, the window is 3.3 hours to give adequate resolution for RUL estimation. The exponential model from equation (2) was fitted to the degradation curves of both cases using the least squares approximation. As a result, the prior estimates of the model parameters (c,b) and the error variance were obtained. The Metropolis-Hastings MCMC algorithm was executed with the prior parameters to estimate the RUL.

\begin{table}[h]
\caption{DEGRADATION GOODNESS METRICS}
\centering
\label{table_3}
\centering
\begin{tabular}{ | p{1.0cm} | p{1.0cm} | p{1.0cm} | p{1.0cm} | p{1.0cm} | p{1.0cm} | p{1.0cm} | p{1.0cm} | p{1.0cm} |  p{1.0cm} |}
\hline
1 & 2 & 3 & 4 & 5 & 6 & 7 & 8 & 9 & 10\\
\hline
0.438 & 0.342 & 0.418 & 0.348 & 0.424 & 0.453 & 0.359 & 0.328 & 0.270 & 0.235\\
\hline
\end{tabular}
\end{table}

Figure \ref{FIG_5} represents RUL estimation with 5\% - 95\% confidence. RUL estimation was done after the last measurement date (the dashed black line). Time intervals between the fault detection time and the last measurement date were approximately two days in both cases. The solid blue line is the estimate of the degradation feature. The estimate was calculated using the exponential model with the model parameters obtained from the RUL algorithm at the last measurement date. The estimation confidence limits were calculated using the model parameters at 5\% - 95\% confidence. The alarm level for the degradation feature (the solid red line) was set to 3.5. The threshold was set higher than the last degradation feature values in both cases for evaluation purposes. The dashed red line represents the last operation date of the bearing. The last operation date was determined when the 5\% confidence limit reached the alarm threshold.

\begin{table}[h]
\caption{FRAMEWORK PERFORMANCE TESTS}
\centering
\label{table_4}
\centering
\begin{tabular}{ | p{3.0cm} | p{2.0cm} | p{2.0cm} | p{2.0cm} | p{2.0cm} | }
\multicolumn{5}{l}{Intel Core i5-6440HQ CPU 2.60GHz - 100 RUNS / FUNCTION}\\
\hline
& FUNC.2 & FUNC.5 & FUNC.10 & FUNC.13\\
\hline
Mean [s] & 0.1616 & 0.0325 & 0.1303 & 24.6400\\
Std [s] & 0.0313 & 0.0258 & 0.0365 & 0.5709\\
\hline
\multicolumn{1}{l}{}\\
\multicolumn{5}{l}{Intel Core i5-4300 CPU 1.90GHz - 100 RUNS / FUNCTION (Fig.\ref{FIG_4})}\\
\hline
& FUNC.2 & FUNC.5 & FUNC.10 & FUNC.13\\
\hline
Mean [s] & 0.2385 & 0.0490 & 0.3256 & 56.1089\\
Std [s] & 0.0137 & 0.0356 & 0.2200 & 2.7115\\
\hline
\end{tabular}
\end{table}

Performance tests for the developed framework were run on two CPUs. Table \ref{table_4} includes averages of 100 runs for the main functions defined in the sequence diagram (Figure \ref{FIG_4}). It is notable that the function call \texttt{getAssetData()} (FUNC.2) also includes the function call \texttt{getMeasurementData()} that reads vibration signal data from external files, which is an operation dependent upon a hard drive. The function call \texttt{estimateRUL()} depends on how many measurements were collected since the bearings started to be used. Other than RUL estimation calculations, execution times of the main functions are not significantly large. However, the performance test should be run on embedded systems that include less computing power.

\section{Conclusions}\label{sec6}

Tribotronic systems are installed in different environments with various configurations. These systems are vulnerable; consequently, complex algorithms are developed to interpret measurements from modern sensors and to make decisions to control the actuators that give feedback to a tribosystem. The fundamental phenomenon of interacting surfaces is the main motivator to build such self-adjusting tribotronic systems.

This paper introduced a unique Tribotronic plug-in software framework that offers assets and data management, feature extraction, fault detection, and RUL estimation for tribotronic systems. The plug-in implementation targets REBs. Further, the plug-in implementation is interchangeable; ergo, it perfectly fits with other tribosystems, such as gears. The framework is platform-independent and also applies to embedded systems. It is extensible and implements functionalities that require considerable amounts of computing power can be implemented in lower-level programming languages. Unit testing capabilities increase the reliability of the implemented plug-in.

The experimental evaluation of the tribotronic plug-in framework were done using bearing vibration data acquired from NASA's prognostics data repository. The evaluation included a run-through from feature extraction to fault detection to RUL estimation. The purpose of the evaluation was not to introduce fault detection or RUL estimation methods, but to show that the framework can handle complex algorithms and produce reliable results. The performance tests demonstrate that the running times are short on ordinary CPUs; however, the lengths of the vibration measurements can be considerably longer, which leads to longer running times. The configuration complexity of the implemented plug-in is low from the point of view of the total number of meta-parameters.

Future research should focus on performance testing of the framework on embedded systems and benchmarking the framework in other tribotronic systems.

\bibliographystyle{unsrt}  
\bibliography{framework}

\end{document}